\documentclass[amssymb,prb,twocolumn,showpacs,superscriptaddress]{revtex4}
\usepackage{graphicx}
\usepackage{epsfig}
\usepackage{longtable}

\begin{document}


\title{Embedded magnetic phases in (Ga,Fe)N: the key role of growth temperature}

\author{A.~Navarro-Quezada}
\affiliation{Institut f\"ur Halbleiter-und-Festk\"orperphysik, Johannes Kepler University, Altenbergerstr. 69, A-4040 Linz, Austria}
\email{andrea.navarro-quezada@jku.at, alberta.bonanni@jku.at}
\author{W.~Stefanowicz}
\affiliation{Laboratory of Magnetism, Bialystok University, ul. Lipowa 41, 15-424 Bialystok, Poland}
\affiliation{Institute of Physics, Polish Academy of Sciences, al. Lotnik\'{o}w 32/46, PL-02-668 Warszawa, Poland}

\author{Tian Li}
\affiliation{Institut f\"ur Halbleiter-und-Festk\"orperphysik, Johannes Kepler University, Altenbergerstr. 69, A-4040 Linz, Austria}

\author{B.~Faina}
\affiliation{Institut f\"ur Halbleiter-und-Festk\"orperphysik, Johannes Kepler University, Altenbergerstr. 69, A-4040 Linz, Austria}

\author{M. Rovezzi}
\affiliation{Institut f\"ur Halbleiter-und-Festk\"orperphysik, Johannes Kepler University, Altenbergerstr. 69, A-4040 Linz, Austria}
\affiliation{Italian Collaborating Research Group, BM08 ``GILDA'' - ESRF, BP 220, F-38043 Grenoble, France}

\author{R.~T.~Lechner}
\affiliation{Institut f\"ur Halbleiter-und-Festk\"orperphysik, Johannes Kepler University, Altenbergerstr. 69, A-4040 Linz, Austria}

\author{T.~Devillers}
\affiliation{Institut f\"ur Halbleiter-und-Festk\"orperphysik, Johannes Kepler University, Altenbergerstr. 69, A-4040 Linz, Austria}

\author{F. d'Acapito}
\affiliation{Italian Collaborating Research Group, BM08 ``GILDA'' - ESRF, BP 220, F-38043 Grenoble, France}
\affiliation{CNR-IOM Sez. di Grenoble, c/o ESRF 6, Rue Jules Horowitz, F-38043 Grenoble (France)}

\author{G.~Bauer}
\affiliation{Institut f\"ur Halbleiter-und-Festk\"orperphysik, Johannes Kepler University, Altenbergerstr. 69, A-4040 Linz, Austria}

\author{M.~Sawicki}
\affiliation{Institute of Physics, Polish Academy of Sciences, al. Lotnik\'{o}w 32/46, PL-02-668 Warszawa, Poland}

\author{T.~Dietl}
\affiliation{Institute of Physics, Polish Academy of Sciences, al. Lotnik\'{o}w 32/46, PL-02-668 Warszawa, Poland}
\affiliation{Institute of Theoretical Physics, University of Warsaw, PL-00-681 Warszawa, Poland}

\author{A.~Bonanni}
\affiliation{Institut f\"ur Halbleiter-und-Festk\"orperphysik, Johannes Kepler University, Altenbergerstr. 69, A-4040 Linz, Austria}
\email{Alberta.Bonanni@jku.at}

\date{\today}

\begin{abstract}
The local chemistry, structure, and magnetism of (Ga,Fe)N nanocomposites
grown by metal organic vapor phase epitaxy is studied by high resolution
synchrotron x-ray diffraction and absorption, transmission electron
microscopy, and superconducting quantum interference device magnetometry as a function of the growth temperature
$T_{\mathrm{g}}$. Three contributions to the magnetization are
identified: i) paramagnetic -- originating from dilute
and non-interacting Fe$^{3+}$ ions substitutional of Ga, and dominating in layers obtained at
the lowest considered $T_{\mathrm{g}}$ (800$^{\circ}$C); ii) superparamagnetic-like -- brought about mainly by ferromagnetic nanocrystals of
$\varepsilon-$Fe$_3$N but also by $\gamma'$-Fe$_4$N and by inclusions of elemental $\alpha$- and
$\gamma$-Fe, and prevalent in films obtained in the
intermediate $T_{\mathrm{g}}$ range; iii) component linear in the magnetic field
and associated with antiferromagnetic interactions -- found to originate
from highly nitridated Fe$_x$N ($x \leq$ 2) phases, like
$\zeta$-Fe$_2$N, and detected in samples deposited at the highest employed
temperature, $T_{\mathrm{g}}$ = 950$^{\circ}$C. Furthermore, depending on
$T_{\mathrm{g}}$, the Fe-rich nanocrystals segregate towards the sample surface
or occupy two-dimensional planes perpendicular to the growth direction.\\
\end{abstract}

\pacs{68.55.Nq, 75.50.Pp, 75.75.-c, 81.05.Ea}

\maketitle

\section{Introduction}
The epitaxy of magnetically doped semiconductors constitutes a versatile mean of fabricating in a self-organized way semiconductor/ferromagnet nanocomposites\cite{Bonanni:2007_SST,Katayama:2007_pssa,Dietl:2008_JAP} with still widely unexplored but remarkable functionalities relevant to spintronics, nanoelectronics, photonics, and plasmonics. In these nanocomposite materials the presence of robust ferromagnetism correlates with the existence of nanoscale volumes containing a large density of magnetic cations, that is with the formation of condensed magnetic semiconductors (CMSs) buried in the host matrix and characterized by a high spin ordering temperature.\cite{Bonanni:2010_CSR} The aggregation of CMSs and, therefore, the ferromagnetism of the resulting composite system shows a dramatic dependence on the growth conditions and co-doping with shallow impurities.

In particular, the understanding and control of CMSs in tetrahedrally coordinated semiconductor films containing transition metals (TMs) -- typical examples being (Ga,Mn)As,\cite{Tanaka:2008_B} (Ga,Mn)N,\cite{Martinez-Criado:2005_APL} (Ge,Mn),\cite{Jamet:2006_NM} (Zn,Cr)Te,\cite{Kuroda:2007_NM} and (Ga,Fe)N \cite{Bonanni:2007_PRB,Bonanni:2008_PRL,Rovezzi:2009_PRB} -- have been lately carefully considered with the necessary aid of nanoscale characterization techniques. Indeed, the control over the CMSs formation as a function of the fabrication parameters, and the possibility to reliably produce on demand CMSs with a predefined size, structure, and distribution in the semiconductor host, are fundamental requisites for the exploitation of these nanostructures in functional devices. At the same time these studies draw us nearer to understand the origin of the ferromagnetic-like features -- persisting up to above room temperature (RT) -- found in a number of semiconductors and oxides. \cite{Liu:2005_JMSME,Coey:2008_JPD}

Dilute zincblende (Ga,Mn)As grown by molecular beam epitaxy (MBE) is known to decompose upon annealing with the formation of embedded MnAs nanocrystals coherent with the GaAs matrix,\cite{De_Boeck:1996_APL} and a striking spin-battery effect produced by these CMSs has been already proven.\cite{Nam-Hai:2009_N}
As an example of the critical role played by growth parameters, MBE Ge$_{1-x}$Mn$_x$ grown below 130$^{\circ}$C is seen to promote the self-assembling
of coherent magnetic Mn-rich nanocolumns whilst a higher growth temperature leads to the formation of hexagonal Ge$_5$Mn$_3$ nanocrystals buried in the Ge host.\cite{Devillers:2007_PRB}

Following theoretical suggestions,\cite{Dietl:2006_NM} it has recently been demonstrated experimentally that it is possible to change the charge state of TM ions in a semiconducting matrix and, therefore, the aggregation energy by co-doping with shallow donors or acceptors.\cite{Kuroda:2007_NM,Bonanni:2008_PRL} In particular, it has been proven that in the model case of wurtzite (wz) (Ga,Fe)N fabricated by metalorganic vapor phase epitaxy (MOVPE) the Fermi-level tuning by co-doping with Mg (acceptor in GaN) or Si (donor in GaN) is instrumental in controlling the magnetic ions aggregation.\cite{Bonanni:2008_PRL} 

The same system has been thoroughly analyzed at the nanoscale by means of advanced electron microscopy as well as by synchrotron-based diffraction and absorption techniques. The structural characteristics have then been related -- together with the growth parameters -- to the magnetic properties of the material as evidenced by superconducting quantum interference device (SQUID) magnetometry.\cite{Bonanni:2007_PRB,Bonanni:2008_PRL,Rovezzi:2009_PRB} It has been concluded that for a concentration of Fe below its optimized solubility limit ($\sim0.4$\% of the magnetic ions) the dilute system is predominantly paramagnetic (PM). For higher concentrations of the magnetic ions (Ga,Fe)N shows either chemical- (intermediate state) or crystallographic-phase separation.~\cite{Bonanni:2007_PRB,Bonanni:2008_PRL,Rovezzi:2009_PRB} In the phase-separated layers a ferromagnetic (FM) behavior persisting far above RT is observed, and has been related to the presence of either Fe-rich regions coherent with the host GaN (in the intermediate state) or of Fe$_{x}$N nanocrystals in the GaN matrix. These investigations appear to elucidate the microscopic origin of the magnetic behavior of (Ga,Fe)N reported by other groups.~\cite{Kuwabara:2001_JJAP,Kane:2006_pssb}

Along the above mentioned lines, in this work we consider further the MOVPE (Ga,Fe)N material system and we reconstruct the phase diagram of the Fe$_{x}$N nanocrystals buried in GaN as a function of the growth temperature. Synchrotron radiation x-ray diffraction (SXRD), extended fine structure x-ray absorption (EXAFS) and x-ray absorption near-edge fine structure (XANES), combined with high-resolution transmission electron microscopy (HRTEM) and SQUID magnetometry allow us to detect and to identify particular Fe$_{x}$N phases in samples fabricated at different growth temperatures $T_{\mathrm{g}}$ as well as to establish a correlation between the existence of the specific phases and the magnetic response of the system. Our results imply, in particular, that self assembled nanocrystals with a high concentration of the magnetic constituent account for ferromagnetic-like features persisting up to above RT. These findings for (Ga,Fe)N do not support, therefore, the recent suggestions that high temperature ferromagnetism of -- the closely related -- oxides is brought about by spin polarization of defects, whereas the role of magnetic impurities is to bring the Fermi energy to an appropriate position in the band gap.~\cite{Coey:2008_JPD} 

We find that already a 5\% variation in the growth temperature is critical for the onset of new Fe$_{x}$N species and we can confirm that an increase in the growth temperature promotes the aggregation of the magnetic ions, resulting in an enhanced density of Fe-rich nanocrystals in the matrix and in a consequent increase of the ferromagnetic response of the system. Moreover, we observe that while in the low-range of growth temperatures the Fe-rich nanoobjects tend to segregate close to the sample surface, at higher $T_{\mathrm{g}}$ two-dimensional assemblies of nanocrystals form in a reproducible way at different depths in the layer, an arrangement expected to have a potential as template for the self-aggregation of metallic nanocolumns.\cite{Fukushima:2006_JJAP} The non-uniform distribution of magnetic aggregates over the film volume here revealed, implies also that the CMS detection may be challenging and, in general, requires a careful examination of the whole layer, including the surface and interfacial regions.

The paper is organized as follows: in the next Section we give a concise summary of the MOVPE process employed to fabricate the (Ga,Fe)N phase-separated samples together with a brief description of the characterization techniques. A table with the relevant samples and relative parameters completes this part.
The central results of this work are reported in Section III and are presented in two sub-sections discussing respectively: i) the detection, identification and structural properties $vs.$ $T_{\mathrm{g}}$ of the different Fe$_{x}$N nanocrystals in phase-separated (Ga,Fe)N, with the distribution of the nanocrystals in the sample volume, and ii) the magnetic properties of the specific families of Fe$_{x}$N phases. In Section IV we sum up the main conclusions and the prospects of this work.

\section{Experimental Procedure}
\subsection{Growth of (Ga,Fe)N}
We summarize here our study by considering a series of wurtzite (Ga,Fe)N samples fabricated by MOVPE in an AIXTRON 200 RF horizontal reactor. All structures have been deposited on $c$-plane sapphire substrates with TMGa (trimethylgallium), NH$_3$, and FeCp$_2$ (ferrocene) as precursors for, respectively, Ga, N and Fe, and with H$_2$ as carrier gas. 

The growth process has been carried out according to a well established procedure,\cite{Simbrunner:2007_APL} namely: substrate nitridation, low temperature (540$^{\circ}$C) deposition of a GaN nucleation layer (NL), annealing of the NL under NH$_3$ until recrystallization and the growth of a $\approx$~1~$\mu$m thick device-quality GaN buffer at 1030$^{\circ}$C. On the GaN buffer, Fe-doped GaN overlayers ($\approx$~700~nm thick) have been deposited at different $T_{\mathrm{g}}$ ranging from 800$^{\circ}$C to 950$^{\circ}$C, with a V/III ratio of 300 [NH$_3$ and TMGa source flow of 1500 standard cubic centimeters per minute (sccm) and 5~sccm, respectively], with an average growth-rate of 0.21 nm/s, and the flow-rate of the Fe-precursor set at 300~sccm. During the whole growth process the samples have been continuously rotated in order to promote the deposition homogeneity, while \textit{in situ} and on line ellipsometry is employed for the real time control over the entire fabrication process. 

The considered samples main parameters, including the Fe concentration, are displayed in Table~\ref{Tab:table1}.

\begin{table} [h]
\begin{ruledtabular}
\caption{\label{Tab:table1} Considered (Ga,Fe)N samples with the corresponding growth temperature, Fe concentration as evaluated by secondary ions mass spectroscopy (SIMS) as well as concentration of the dilute paramagnetic Fe$^{3+}$ ions $x_{\mathrm{Fe}^{3+}}$ and a lower limit of the concentration of Fe ions $x_{\mathrm{Fe}_N}$ contributing to the Fe-rich nanocrystals, as obtained from magnetization data.}
\label{Tab:table1}
\begin{tabular}{|c|cccc|}
Sample & $T_{\mathrm{g}}$ & Fe concentration    & $x_{\mathrm{Fe}^{3+}}$ & $x_{\mathrm{Fe}_N}$ \\
 &   $^{\circ}$C       & [$10^{20}$ cm$^{-3}$]        & [$10^{19}$ cm$^{-3}$] & [$10^{19}$ cm$^{-3}$] \\
\hline
S690 & 800 & 1 & $3.2$ & $0.1$\\
S687 & 850 & 2 & $2.9$ & $1.7$ \\
S680 & 850 & 2 & $2.7$ & $1.5$\\
S987 & 900 & 4 & $2.4$ & $1.6$ \\
S691 & 950 & 4 & $2.9$ & $3.2$ \\
\end{tabular}
\end{ruledtabular}
\end{table}

\subsection{Synchrotron x-ray diffraction -- experimental}
 Coplanar SXRD measurements have been carried out at the Rossendorf Beamline BM20 of the European Synchrotron Radiation Facility (ESRF) in Grenoble -- France, using a photon energy of 10.005~keV. The x-ray data correspond to the diffracted intensities in reciprocal space along the sample surface normals.
The beamline is equipped with a double-crystal Si(111) monochromator with two collimating/focusing mirrors (Si and Pt-coating) for rejection of higher harmonics, allowing measurements in an energy range of 6 to 33~keV. The symmetric $\omega=2\theta$ scans are acquired using a heavy-duty 6-circle Huber diffractometer and the most intense peaks are found for 2$\theta$ up to 40$^{\circ}$.

\subsection{XAFS -- experimental and method}\label{sec:xafs_exp}
 X-ray absorption fine structure (XAFS) measurements at the Fe K edge (7112~eV) are carried out at the ``GILDA'' Italian collaborating research group beam-line~\cite{d'Acapito:1998_EN} (BM08) of ESRF under the same experimental conditions reported in Ref.~\onlinecite{Rovezzi:2009_PRB}, collecting both the XANES and EXAFS spectra, and employing the following method for the analysis. 
 
A set of model compounds is established: Fe substitutional of Ga in GaN (Fe$_{\rm Ga}$),~\cite{Rovezzi:2009_PRB} $\zeta$-Fe$_2$N,~\cite{Rechenbach:1996_JAC} $\varepsilon$-Fe$_3$N,~\cite{Jacobs:1995_JAC} $\gamma$-Fe$_4$N,~\cite{Jacobs:1995_JAC} $\alpha$-Fe~\cite{Swanson1955} and $\gamma$-Fe.~\cite{Gorton1965} For these input structures the XANES absorption spectra are calculated using the {\sc fdmnes} code~\cite{Joly:2009_JPC} while the EXAFS scattering expansion signals are computed with the {\sc feff8.4} code~\cite{Ankudinov:1998_PRB} in order to de-correlate the structural results to a specific software choice. In both cases muffin-tin potentials and the Hedin-Lunqvist approximation for their energy-dependent part is used, with self-consistent potential calculation for enhancing the accuracy in the determination of the Fermi energy ($E_{\rm F}$). 

X-ray polarization is taken into account for Fe$_{\rm Ga}$ while unpolarized simulations are conducted for the other phases assuming a random orientation of the nanocrystals in the sample. In addition, for XANES the convergence of the results is tested against the increasing in the input cluster size ($>$150 atoms) and the method is validated by experimental values from Fe$_{\rm Ga}$ and $\alpha$-Fe. The resulting simulated spectra are then convoluted $\textit{via}$ an energy-dependent function as implemented in {\sc fdmnes}~\cite{Joly:2009_JPC} plus a Gaussian experimental broadening of 1~eV and fitted to the normalized XANES experimental data in the energy range from -20 to 80~eV relative to $E_{\rm F}$ with a linear combination analysis using the {\sc Athena} graphical interface~\cite{Ravel:2005_JSR} to {\sc i\-feffit}.~\cite{Newville:2001_JSR} 

All the possible combinations with a maximum of three spectra per fit (maximum of six fit parameters: amplitude and energy shift) are tested and the best fit is chosen on the basis of the $\chi^2$ statistics, discarding unphysical results. Finally, the XANES results are independently checked through the quantitative analysis of the EXAFS data where the background-subtracted (via the {\sc viper} program\cite{Klementev:2001_JPDAP}) $k^2$-weighted fine-structure oscillations, $\chi(k)$, are fitted in the Fourier-transformed space.

\subsection{High resolution transmission electron microscopy -- experimental}
 The HRTEM studies are performed on cross-sectional samples prepared by standard mechanical polishing followed by Ar$^{+}$ ion milling at 4~kV for about 1~h. Conventional diffraction contrast images in bright-field imaging mode and high-resolution phase contrast pictures were obtained from a JEOL 2011 Fast TEM microscope operating at 200~kV and capable of an ultimate point-to-point resolution of 0.19~nm and allowing to image lattice fringes with a 0.14~nm resolution. 
 
Additionally, energy dispersive x-ray (EDS) analysis has been performed \textit{via} an Oxford Inca EDS equipped with a silicon detector to obtain information on the local composition. Selected area electron diffraction (SAED) and fast Fourier transform (FFT) procedures are employed to study scattering orders and $d$-spacing for respectively the larger and the smaller nanocrystals. 

\subsection{SQUID magnetometry -- experimental}
 The magnetic properties have been investigated in a
Quantum Design MPMS XL 5 SQUID magnetometer between 1.85 and 400~K
and up to 50~kOe following the methodology described
previously.\cite{Stefanowicz:2010_PRB} 

The difference between the
magnetization values measured up to 50~kOe at 1.8~K and 5~K is
employed to determine the concentration $x_{\text{Fe$^{3+}$}}$ of
paramagnetic Fe$^{3+}$ ions in the layers.\cite{Pacuski:2008_PRL}
The lower limit of the concentration of Fe ions contributing to
the Fe-rich nanocrystals, as well as an assessment of their Curie
temperature is inferred from magnetization curves at higher
temperatures. 

Finally, measurements of field cooled (FC) and zero field
cooled (ZFC) magnetization hint to the influence of the growth
temperature on the size distribution of the nanocrystals.

\section{Results}
\subsection{Fe$_{x}$N phases \textit{vs.} $T_{\mathrm{g}}$ in crystallographically separated (Ga,Fe)N}

 Before entering into the detailed discussion of our studies, we would like to point out that the reproducibility of the 
data has been accurately tested and: i) different samples grown under the same conditions have been characterized, ii) all measurements (SXRD, HRTEM, etc...) have been repeated in different runs on the same samples and we can conclude that both the (Ga,Fe)N structures are stable over time and the formation of different phases is reproduced when the growth conditions are fidely replicated. 

\begin{figure}
    \includegraphics[width=\columnwidth]{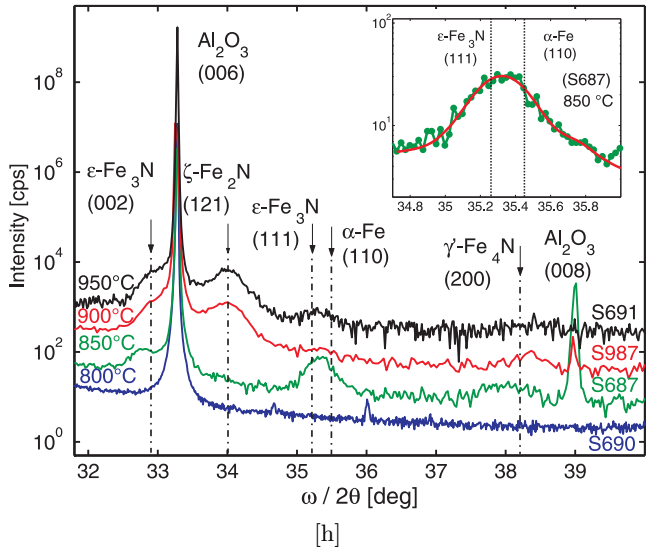} [h]
    \caption{(Color online) SXRD spectra for (Ga,Fe)N layers deposited at different growth temperatures. Inset: peak at 35.3$^{\circ}$ deconvoluted into two components assigned to diffraction maxima (111) of $\varepsilon$-Fe$_3$N  and (110) of $\alpha$-Fe  [experiment (dotted line) and fit (smooth line)].}
    \label{fig:SXRD}
\end{figure}

\begin{figure} 
    \includegraphics[width=0.98\columnwidth]{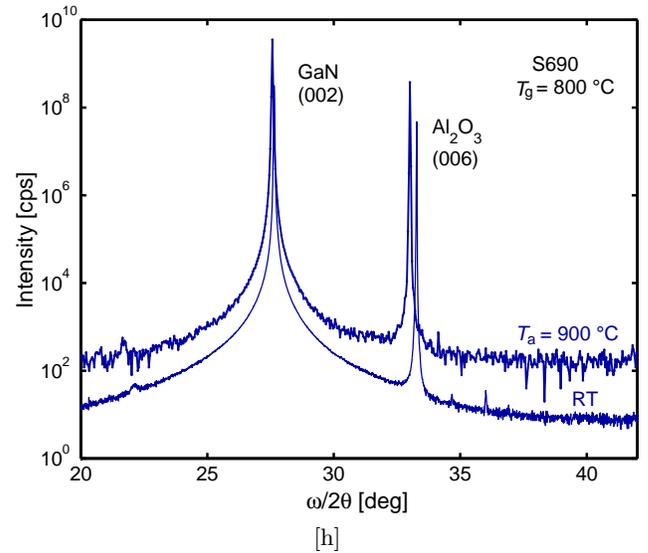} [h]
    \caption{SXRD spectra for a dilute (Ga,Fe)N sample (S690) as grown and upon $\textit{in situ}$ annealing at $T_{\mathrm{a}}$ = 900$^{\circ}$C for 1~h, indicating that post-growth annealing does not induce - in the SXRD sensitivity range - crystallographic decomposition.}
    \label{fig:SXRD_ann}
\end{figure}

In Fig.~\ref{fig:SXRD} we report SXRD diffraction spectra for the (Ga,Fe)N samples  grown at different temperatures, as listed in Table I. For the layer S690 fabricated at 800$^{\circ}$C we have no evidence of secondary phases and only diffraction peaks originating from the sapphire substrate and from the GaN matrix are revealed, in agreement with HRTEM measurements showing no phase separation. Moreover, in order to test the stability of the dilute phase, we have annealed the samples up to $T_{\mathrm{a}}$ = 900$^{\circ}$C and $\textit{in situ}$ SXRD measurements upon annealing do not reveal the onset of any secondary phases, as reported in Fig.~\ref{fig:SXRD_ann}, in accord with the behavior of dilute Mn in GaN \cite{Stefanowicz:2010_PRB} and in contrast with (Ga,Mn)As where post-growth annealing is found to promote the precipitation of MnAs nanocrystals. \cite{Tanaka:2008_B}

Moving to a $T_{\mathrm{g}}$ of 850$^{\circ}$C (S687) different diffraction peaks belonging to secondary phases become evident. We have previously reported\cite{Bonanni:2007_PRB} that when growing (Ga,Fe)N at this temperature, one dominant Fe-rich phase is formed, namely wurzite $\varepsilon$-Fe$_3$N, for which we identify two main peaks corresponding to the (002) and the (111) reflexes, respectively. A closer inspection of the (111)-related feature and a fit with two gaussian curves centered at 35.2$^{\circ}$ and 35.4$^{\circ}$, gives evidence of the presence of the (110) reflex from cubic metallic $\alpha$-Fe. Moreover, the broad feature appearing around 38$^{\circ}$ is associated to the (200) reflex of face centered cubic ($fcc$) $\gamma$'-Fe$_4$N, that crystallizes in an inverse perovskite structure.\cite{Jack:1952_AC} From the position of the peak, we can estimate that these nanocrystals are strained.

\begin{figure} [htb]
    \includegraphics[width=0.93\columnwidth]{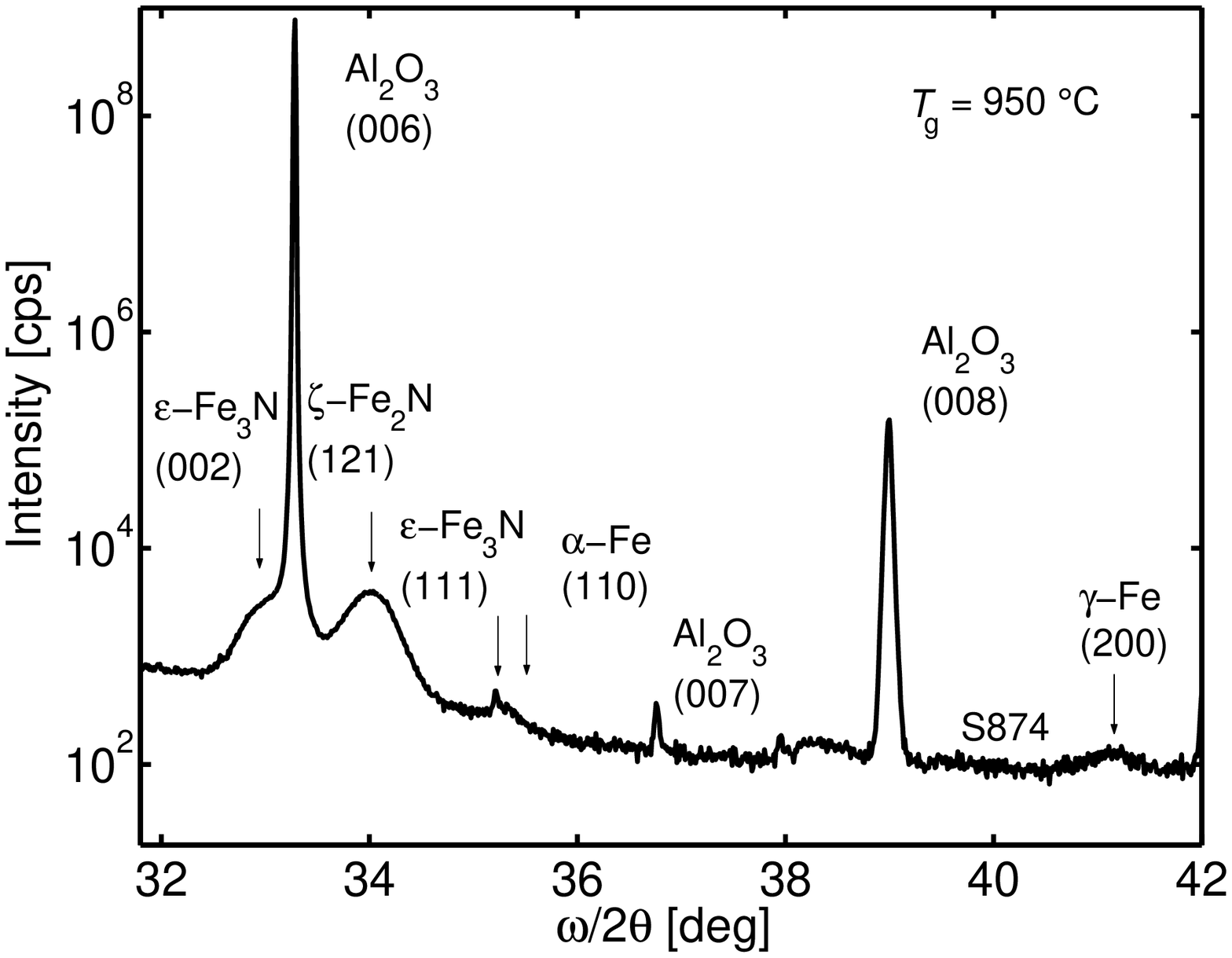} [h]
    \caption{SXRD for a sample (S874) grown at 950$^{\circ}$C evidencing the aggregation of (200) $\gamma$-Fe in the (Ga,Fe)N layer.}
    \label{fig:SXRD_950}
\end{figure}

As the growth temperature is increased to 900$^{\circ}$C (S987) there is no contribution left from the (110) $\alpha$-Fe phase, and the signal from the (111) of $\varepsilon$-Fe$_3$N is significantly quenched, indicating the reduction in either size or density of the specific phase. Furthermore, an intense peak is seen at 34$^{\circ}$, corresponding to the (121) contribution from orthorhombic $\zeta$-Fe$_2$N. This phase crystallizes in the $\alpha$-PbO$_2$-like structure, where the Fe atoms show a slightly distorted hexagonal close packing (\textit{hcp}), also found for $\varepsilon$-Fe$_3$N.\cite{Jacobs:1995_JAC} 

The structural resemblance of $\varepsilon$-Fe$_3$N and the $\zeta$-Fe$_2$N is remarkable, as the \textit{hcp} arrangement in $\varepsilon$-Fe$_3$N is nearly retained in $\zeta$-Fe$_2$N.\cite{Rechenbach:1996_JAC} This gives a hint of the likely direct conversion of phase from $\varepsilon$-Fe$_3$N into $\zeta$-Fe$_2$N. The diffraction peak from (200) $\gamma$'-Fe$_4$N is still present at this temperature, but its position is slightly shifted to its bulk value. A similar behavior is observed for the diffraction from (200) $\varepsilon$-Fe$_3$N (002), shifted from 32.78$^{\circ}$ to 32.9$^{\circ}$.

At a growth temperature of 950$^{\circ}$C (S691) the diffraction peak of (200) $\gamma$'-Fe$_4$N recedes, indicating the decomposition of this \textit{fcc} phase at temperatures above 900$^{\circ}$C, in agreement with the phase diagram for free standing Fe$_x$N,\cite{Jacobs:1995_JAC} reporting cubic $\gamma$'-Fe$_4$N as stable at low temperatures. Only the (002) $\varepsilon$-Fe$_3$N- and the (121) $\zeta$-Fe$_2$N-related diffraction peaks are preserved with a constant intensity and position with increasing temperature, suggesting that at high $T_{\mathrm{g}}$ these two phases and their corresponding orientations, are noticeably stable. Furthermore, in samples grown at this $T_{\mathrm{g}}$ the peak from (200) $\gamma$-Fe is detected around 41.12$^{\circ}$, as reported in Fig.~\ref{fig:SXRD_950}, in agreement with the XAFS data discussed later in this Section.

\begin{figure}[htb]
    \centering
        \includegraphics[width=1.1\columnwidth]{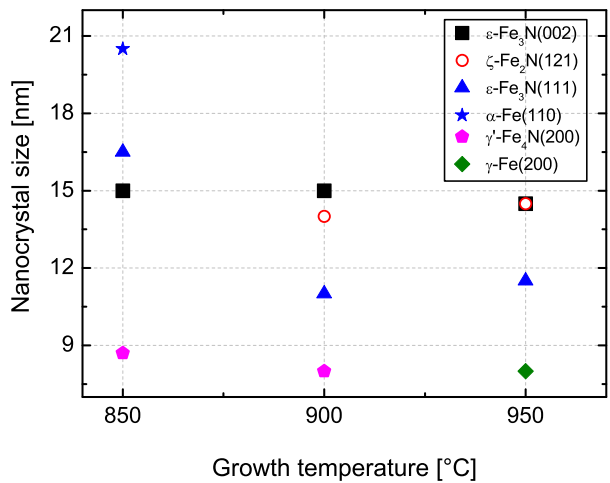}
    \caption{(Color online) Average size $vs.$ $T_{\mathrm{g}}$ of nanocrystals in the different Fe$_x$N phases, as determined from SXRD.}
    \label{fig:nanosize}
\end{figure}

Following the procedure employed previously,\cite{Lechner:2009_APL} and based on the Williamson-Hall formula method,\cite{Williamson:1953_AcMetal} we obtain the approximate average nanocrystals size from the full-width at half maximum (FWHM) of the diffraction peaks in the radial ($\omega/2\theta$) scans. The FWHM of the (002) $\varepsilon$-Fe$_3$N, of the (200) $\gamma$'-Fe$_4$N, and of the (121) $\zeta$-Fe$_2$N diffraction peaks are comparable for samples grown at different temperatures, indicating that the average size of the corresponding nanocrystals is also constant, as summarized in Fig.~\ref{fig:nanosize}. 

\begin{figure}[htb]
    \centering
        \includegraphics[width=0.8\columnwidth]{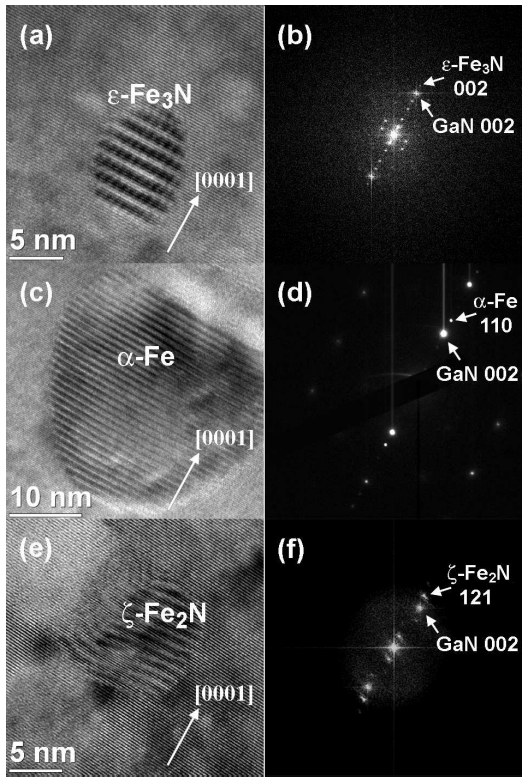}
    \caption{Transmission electron micrographs of different Fe$_x$N phases: a) HRTEM image of a $\varepsilon$-Fe$_{3}$N nanocrystal; (b) the corresponding FFT image, revealing that the \textit{d}-spacing along the growth direction is about 0.216~nm. (c) HRTEM image on $\alpha$-Fe nanocrystal in sample S687; (d) SAED pattern on the enclosing area. (e) HRTEM image of a $\zeta$-Fe$_{2}$N nanocrystal; (f) the corresponding FFT image, revealing that the \textit{d}-spacing along the growth direction is about 0.211~nm.}
    \label{fig:TEM}
\end{figure}

The (111) $\varepsilon$-Fe$_3$N signal intensity is seen to change abruptly when comparing the results for the sample grown at 850$^{\circ}$C to those from the layers fabricated at higher temperatures. From the FWHM for this particular orientation we can estimate that the nanocrystal average size adjusts between 16.5 and 12.0~nm in the considered temperature range. At high temperatures, the size then remains constant up to 950$^{\circ}$C. The size of the $\alpha$-Fe nanocrystals can only be estimated for the sample grown at 850$^{\circ}$C, where the corresponding diffraction peak can easily be resolved and suggests an average size of these objects larger than that of the other identified phases, as confirmed by the HRTEM images reported in Fig.~\ref{fig:TEM}.

\begin{figure}[h]
  \includegraphics[width=0.5\textwidth]{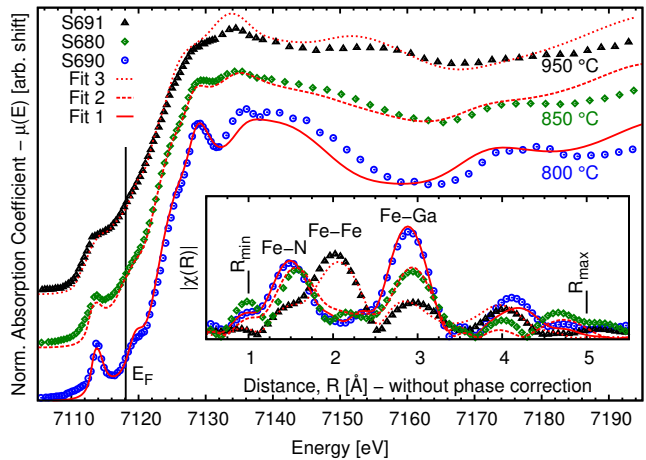}
  \caption{(Color online) Normalized XANES spectra (main plot) and the amplitude of the Fourier-transforms (inset) of the $k^2$-weighted EXAFS in the range from 2.5 to 10.0 ~\AA$^{-1}$ for three samples (points) grown at different temperatures, with their relative fits (lines) summarized in Table~\ref{tab:xafs}.}
  \label{fig:xafs}
\end{figure}

The XAFS study on the (Ga,Fe)N samples fabricated at different $T_{\mathrm{g}}$ permits to have a structural description of the atomic environment around the absorbing species from a local point of view, complementary to SXRD. The experimental data are reported in Fig.~\ref{fig:xafs} with the relative fits obtained by following the method described in Sec.~\ref{sec:xafs_exp}. Qualitatively an evolution with $T_{\mathrm{g}}$ is visible and it is quantitatively confirmed by the results summarized in Table~\ref{tab:xafs}. 

In particular, from the XANES analysis -- sensitive to the occupation site symmetry and multiple scattering effects -- it is possible to infer how the composition of the different phases evolves with increasing $T_{\mathrm{g}}$: Fe$_{\rm Ga}$ reduces, while $\varepsilon$-Fe$_3$N increases up to 950$^\circ$C when the precipitation is in favor of $\zeta$-Fe$_2$N and $\gamma$-Fe. This behavior is confirmed also by the EXAFS spectra given in the inset to Fig.~\ref{fig:xafs}, where the first three main peaks in the fit range from $R_{\rm min}$ to $R_{\rm max}$ represent, respectively, the average Fe-N, Fe-Fe and Fe-Ga coordination. 

In addition, the signal present at longer distances confirms the high crystallinity and permits to include important multiple scattering paths in the fits for a better identification of the correct phase. In fact, from the Fe-Fe distances it is possible to distinguish Fe$_x$N ($\approx$~2.75~\AA) from pure Fe phases ($\approx$~2.57~\AA), while the distinction between $\alpha$-Fe and $\gamma$-Fe is possible with the different multiple scattering paths generated from the body centered cubic ($bcc$) and from the $fcc$ structure, respectively.

\begin{table*}[htbp]
\begin{ruledtabular}
  \caption{Quantitative results of the XAFS analysis (best fits). XANES: composition ($x$) and energy shift relative to $E_{\rm F}$ ($\Delta$$E$) for each structure; EXAFS: average distance ($R$) and Debye-Waller factor ($\sigma^2$) for the first three coordination shells around the absorber. For each phase the coordination numbers are kept to the crystallographic ones and rescaled by the relative fractions found by XANES and a global amplitude reduction factor, $S_0^2$, of 0.93(5) as found for Fe$_{\rm Ga}$. Error bars on the last digit are reported in parentheses.}
  \label{tab:xafs}
    \begin{tabular}{|c|cccccccc|cccccc|}
      Fit & \multicolumn{8}{c|}{XANES} & \multicolumn{6}{c|}{EXAFS} \\
      & \multicolumn{2}{c}{Fe$_{\rm Ga}$} & \multicolumn{2}{c}{$\zeta$-Fe$_2$N} & \multicolumn{2}{c}{$\varepsilon$-Fe$_3$N} & \multicolumn{2}{c|}{$\gamma$-Fe} & \multicolumn{2}{c}{Fe-N} & \multicolumn{2}{c}{Fe-Fe} & \multicolumn{2}{c|}{Fe-Ga}\\
      & $x$ & $\Delta$$E$ & $x$ & $\Delta$$E$ & $x$ & $\Delta$$E$ & $x$ & $\Delta$$E$ & $R$    & $\sigma^2$ & $R$ & $\sigma^2$ & $R$ & $\sigma^2$\\
      & & (eV) & & (eV) & & (eV) & & (eV) & (\AA) & (10$^{-3}$\AA$^2$) & (\AA) & (10$^{-3}$\AA$^2$) & (\AA) & (10$^{-3}$\AA$^2$)\\
      \hline
      1 & 0.9(1) & 1.3(5) & -      & -      & 0.1(1) & 2.9(9) &     -  &      -  & 1.99(1) & 5(2)  & 2.75(5) & 13(5) & 3.20(1) & 7(1)\\ 
      2 & 0.6(1) & 1.1(5) & -      & -      & 0.4(1) & 1.8(5) &     -  &      -  & 2.00(2) & 4(1)  & 2.76(2) & 9(4)  & 3.20(1) & 8(1)\\ 
      3 & 0.2(1) & 1.0(5) & 0.4(1) & 4.5(5) &    -   &     -  & 0.4(1) & -0.3(5) & 1.95(4) & 10(9) & 2.60(5) & 15(9) & 3.18(2) & 4(2)\\ 
    \end{tabular}
    \end{ruledtabular}
\end{table*}

The presence of the different Fe$_x$N phases detected with SXRD has been confirmed also by HRTEM measurements on the considered samples, as reported in Fig.~\ref{fig:TEM}. All the HRTEM images presented here have been taken along the $[10\overline{1}0]$ zone axis. 

\begin{figure}
    \centering
        \includegraphics[width=0.8\columnwidth]{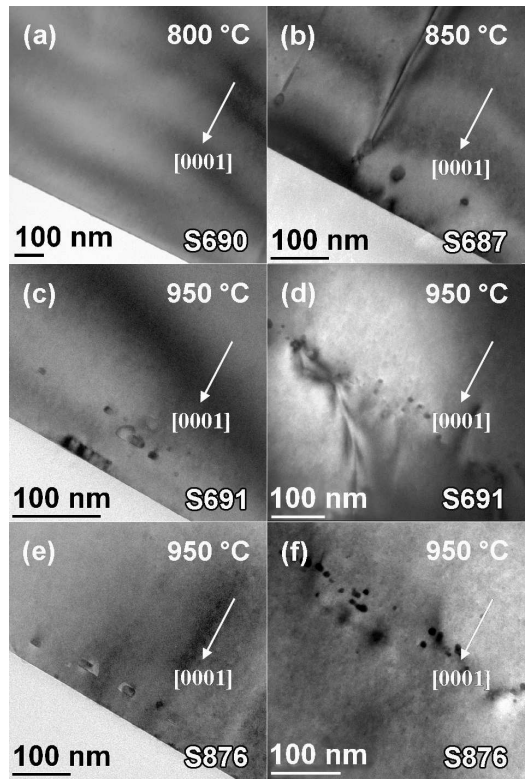} 
    \caption{(Color online) TEM images: distribution of the Fe-rich nanocrystals with increasing growth temperature. (a) $T_{\mathrm{g}}$ = 800$^{\circ}$C (S690) -- dilute (Ga,Fe)N; (b) $T_{\mathrm{g}}$ = 850$^{\circ}$C (S687), Fe-rich nanocrystals concentrated in proximity of the samples interface solely; $T_{\mathrm{g}}$ = 950$^{\circ}$C (S691) -- Fe-rich nanocrystals segregating in proximity of the sample surface (c) and of the interface between the GaN buffer and the Fe-doped layer (d); (e),(f) $T_{\mathrm{g}}$ = 950$^{\circ}$C (S876) -- to be compared with the TEM images of S691 in (c) and (d): reproducibility in the distribution of the Fe-rich nanocrystals for different samples grown at the same $T_{\mathrm{g}}$.}
    \label{fig:distribution}
\end{figure}

By using the SAED technique for the larger nanocrystals and a FFT combined with a subsequent reconstruction for the smaller objects, we have studied the foreign scattering orders and the \textit{d}-spacings along the growth direction. As shown in Fig.~\ref{fig:TEM}(a), the transitional Moir\'e fringes indicate that there is a set of planes parallel to the GaN (002) ones with a similar \textit{d}-spacing inside the nanocrystal. The corresponding FFT image shown in Fig.~\ref{fig:TEM}(b) gives an additional diffraction spot close to GaN (002), corresponding to a \textit{d}-spacing of 0.217~nm, matching the $\textit{d}_{002}$ of $\varepsilon$-Fe$_{3}$N. The $\varepsilon$-Fe$_{3}$N phase is found in all the considered samples, with the exception of the one grown at 800$^{\circ}$C (S690, dilute). The phase $\varepsilon$-Fe$_{3}$N has the closest structure to wurtzite GaN and we can assume that the formation of $\varepsilon$-Fe$_{3}$N is, thus, energetically favored.\cite{Li:2008_JCG}

The micrograph displayed in Fig.~\ref{fig:TEM}(c) has been obtained from the layer grown at 850$^{\circ}$C and refers to a nanocrystal located in the proximity of the sample surface. The corresponding SAED pattern in Fig.~\ref{fig:TEM}(d) reveals that the \textit{d}-spacing of the lattice planes overlapping the GaN matrix has a value of 0.203~nm, matching the $\textit{d}_{110}$ of $\alpha$-Fe. For values of $T_{\mathrm{g}}$ between 900 and 950$^{\circ}$C, nanocrystals like the one represented in Fig.~\ref{fig:TEM}(e) are found. The FFT image shown in Fig.~\ref{fig:TEM}(f), reveals that the additional \textit{d}-spacing is 0.211~nm, corresponding to the $\textit{d}_{121}$ of $\zeta$-Fe$_{2}$N.

\begin{figure}
    \centering
        \includegraphics[width=1.0\columnwidth]{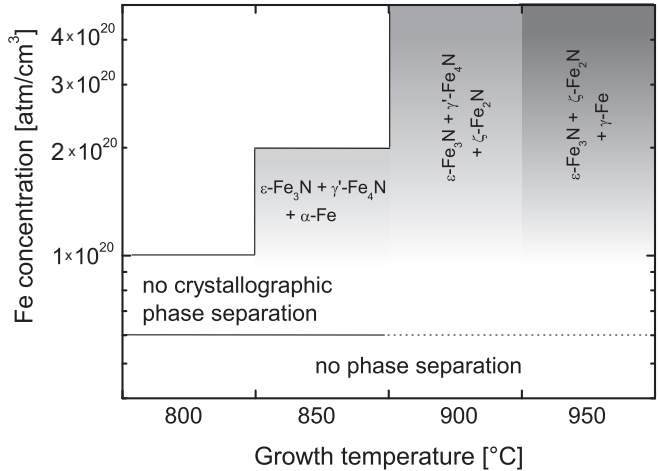}
    \caption{A phase diagram of (Ga,Fe)N as a function of the growth temperature.}
    \label{fig:phasediagram}
\end{figure}

It should be underlined here that the size of the nanocrystals in the HRTEM images is smaller than the average value obtained from SXRD. This discrepancy originates from the fact that a cross-sectional TEM specimen must be rendered considerably thin, in order to achieve a sufficient transparency to the electron beam. Therefore, a nanocrystal is usually only partly enclosed in the investigated area. At the same time, in low magnification micrographs on thicker volumes the size of the objects becomes comparable to the average value determined by the SXRD studies.

Cross-sectional low-magnification TEM measurements permit to observe that while at the lower growth temperatures the Fe-rich nanoobjects tend to segregate close to the sample surface as seen in Fig.~\ref{fig:distribution}(b), at higher $T_{\mathrm{g}}$ two-dimensional assemblies of nanocrystals form in a reproducible way -- as proven by comparing Fig.~\ref{fig:distribution}(c),(d) and Fig.~\ref{fig:distribution}(e),(f) -- and this arrangement is expected to be instrumental as template for the self-aggregation of metallic nanocolumns.\cite{Fukushima:2006_JJAP} 

Summarizing the SXRD, XAFS and HRTEM findings, a phase diagram of the Fe-rich phases formed in (Ga,Fe)N as a function of the growth temperature is constructed and reported in Fig.~\ref{fig:phasediagram}, showing the dominant phases for each temperature interval.

According to Ref.~\onlinecite{Jack:1952_AC} when the concentration of the interstitial-atoms in the $\varepsilon$ phase is increased by only 0.05 atoms/100 Fe, a phase transition from $\varepsilon$ to $\zeta$ occurs. In this process the Fe atoms retain their relative positions but there is a slight anisotropic distortion of the $\varepsilon$ lattice that reduces the symmetry of the (nano)crystal to $\zeta$-orthorhombic. The hexagonal unit cell parameter a$_{hex}$ of $\varepsilon$-Fe$_3$N splits into the parameters $b_{\mathrm{orth}}$ and $c_{\mathrm{orth}}$ in $\zeta$-Fe$_2$N. Moreover, according to the Fe \textit{vs.} N phase diagram the orthorhombic phase contains a higher percentage of nitrogen\cite{Jack:1952_AC} compared to the hexagonal one, and this guides us to conjecture that the higher the growth temperature, the more nitrogen is introduced into the system.

Remarkable is the fact that by increasing the growth temperature the (002) $\varepsilon$-Fe$_3$N is preserved, while the (111) oriented nanocrystals are not detected. A focused study would be necessary to clarify the kinetic processes taking place between 850$^{\circ}$C and 900$^{\circ}$C. Moreover, it is still to be clarified whether the fact that the $\varepsilon$-Fe$_3$N nanocrystals oriented along the growth direction are stable, while the ones lying out of the growth plane are not, may be related to differences in surface energy.

The Fe$_x$N phases found in our (Ga,Fe)N samples are listed in Table~\ref{tab:table2}, together with their crystallographic structure, lattice parameters, $d$-spacing for the diffracted peaks, and magnetic properties.

\begin{table*} 
\begin{ruledtabular}
\caption{\label{tab:table2}Structural and magnetic parameters of
the Fe-rich phases found in the considered (Ga,Fe)N samples.}
\label{tab:table2}
\begin{tabular}{|c|cccc|cccc|}

&&\multicolumn{3}{c|}{Lattice parameter\cite{Eck:1999_JCM}}&\multicolumn{3}{c}{$d$-spacing}&\\
&&\multicolumn{3}{c|}{-------------------------------}&\multicolumn{3}{c}{-------------------------------}&\\
& Structure & $a$(nm)& $b$(nm)& $c$(nm)& literature value\cite{icdd:2009}& SXRD& HRTEM& $\mu_{B}$\\
\hline
$\gamma'$-Fe$_4$N & $\textit{fcc}$ & 0.382 & -- & -- & 0.189 & 0.188-0.189 & 0.188 & 2.21~\cite{Eck:1999_JCM}\\
$\varepsilon$-Fe$_3$N & wz & 0.469 & -- & 0.438 & 0.2189$_{(002)}$ & 0.2188$_{(002)}$ & 0.2178$_{(002)}$ & 2.0~\cite{Leineweber:1999_JAC} \\
& & & & & 0.208$_{(111)}$ & 0.206$_{(111)}$ & --  &\\
$\zeta$-Fe$_2$N & ortho & 0.443 & 0.554 & 0.484 & 0.2113 & 0.2114 & 0.211 & 1.5~\cite{Eck:1999_JCM}\\
$\alpha$-Fe& $\textit{bcc}$ & 0.286 & -- & -- & 0.202 & 0.204 & 0.203 & 2.2~\cite{Keavney:1995_PRL}  \\
$\gamma$-Fe& $\textit{fcc}$  & 0.361 & -- & -- & 0.180$_{(200)}$ & 0.176$_{(200)}$ & --    & 0.3--1.6~\cite{Shi:1996_PRB} \\ 
& & & & & 0.210$_{(111)}$ & -- & -- &\\

\end{tabular}
\end{ruledtabular}
\end{table*}

Further focused studies are required in order to clarify the kinetic mechanisms of segregation and possibly the range of parameters that could allow the selectivity of the species in different two-dimensional regions of the doped layers. 

\subsection{Magnetic properties of Fe$_x$N phases}

As reported in Table~\ref{tab:table2}, the different Fe$_x$N phases we identify in the considered samples are expected to show specific magnetic responses. The $\varepsilon$-Fe$_3$N phase, predominant in the samples grown at 850$^{\circ}$C, is ferromagnetic with a Curie temperature $T_{\mathrm{C}}$ of 575~K.\cite{Leineweber:1999_JAC} 

The $\gamma$'-Fe$_4$N phase, also present though in lesser amount in these layers, is FM too, with a $T_{\mathrm{C}}$ of 750~K.\cite{Jack:1952_AC} For the samples deposited at temperatures above 850$^{\circ}$C, the dominant and stable phase becomes $\zeta$-Fe$_2$N.

The magnetic response of these (Ga,Fe)N layers is quite typical for semiconductors containing TM ions at concentration above or close to the solubility limits. Regardless of the prevailing diamagnetic component from the sapphire substrate -- that we compensate with the procedure detailed elsewhere  \cite{Stefanowicz:2010_PRB} -- the field dependency of  magnetization $M(H)$ is characterized primarily by a dominant paramagnetic contribution at low temperatures from diluted substitutional Fe$^{3+}$ ions and by a superparamagnetic-like component saturating (relatively) fast and originating from various magnetically ordered nanocrystals with high Fe content. Among them, the FM hexagonal $\varepsilon$-Fe$_{3-x}$N have the highest density, according to the SXRD and HRTEM studies discussed above. Despite the richness of different phases, it is relatively straightforward to separate these major components and to treat them -- to a large extent -- qualitatively.

\begin{figure*}[t]
   \begin{center}
        \includegraphics[width=1.98\columnwidth]{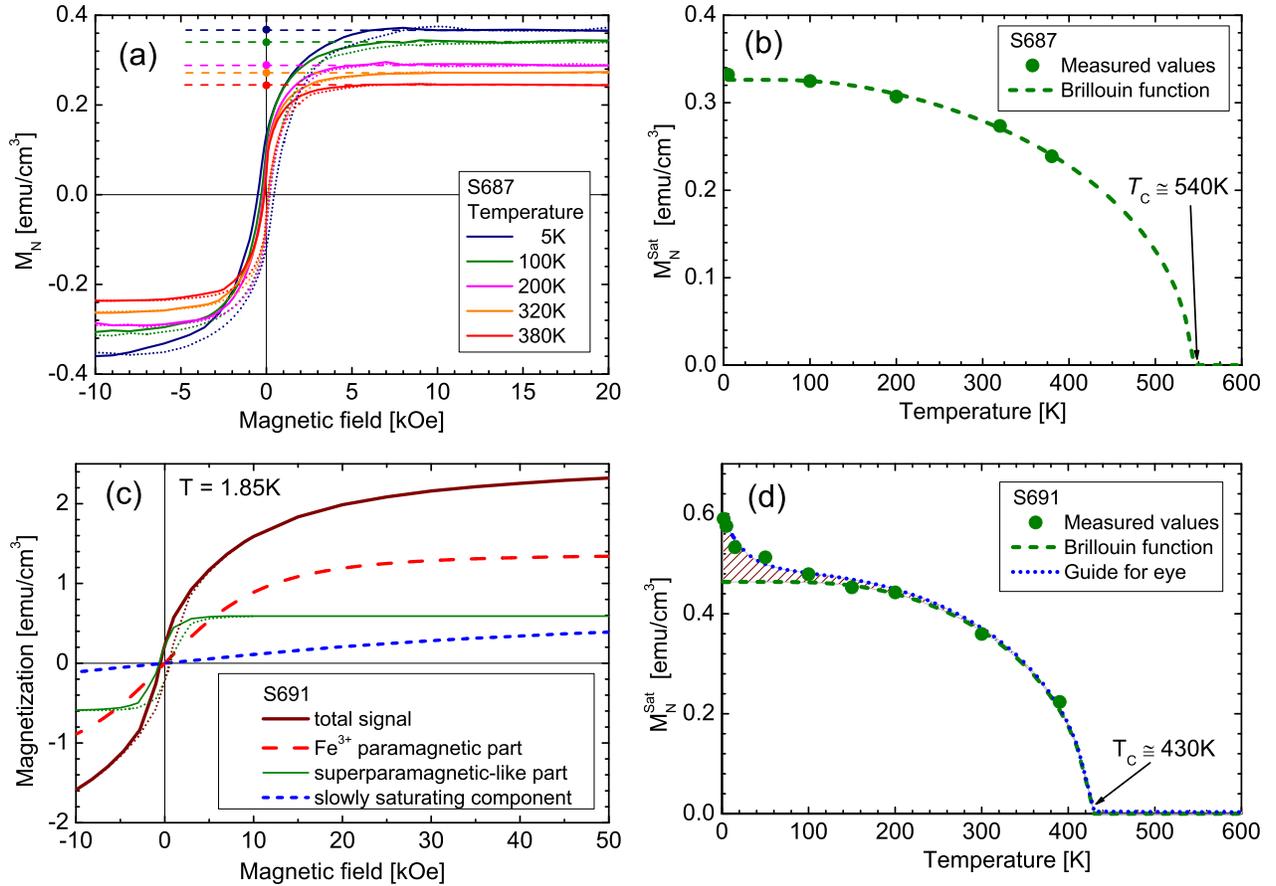}
   \end{center}
\caption{(Color online) (a) Magnetic field dependence of the
nanocrystals magnetization $M_{\mathrm{N}}$ at selected temperatures for
sample S687 ($T_{\mathrm{g}}$ = 850$^{\circ}$C). Each $M_{\mathrm{N}}(H)$
curve has been measured from the maximum positive to the maximum negative
field, only and the dotted lines obtained by numerical invertion are guides
for the eye. The dashed lines represent the saturation level of
magnetization at each temperature.
(b) Bullets - temperature dependence of the saturation
magnetization $M_{\mathrm{N}}^{\mathrm{Sat}}$ obtained from panel (a).
Dashed line - the Brillouin function for the magnetic moment of
2$\mu_{\mathrm{B}}$ per Fe atom.
(c) Three major contributions to the total magnetic signal (thick
brown solid) for sample S691 ($T_{\mathrm{g}}$ = 950$^{\circ}$C): i)
paramagnetic from Fe$^{3+}$ (thick red dashed), ii) high-$T_{\mathrm{C}}$
superparamagnetic-like (thin green solid) from the nanocrystals and iii) slowly
saturating component (blue short dashed). Also here only a half of the
full hysteresis loop was measured and the dotted lines obtained by
numerical reflection are guides for the eye. (d) Bullets - temperature
dependence of $M_{\mathrm{N}}^{\mathrm{Sat}}$ for sample S691. Dashed line
- the Brillouin function for the magnetic moment of 2$\mu_{\mathrm{B}}$ per
atom Fe. The blue dotted line follows the excess of
$M_{\mathrm{N}}^{\mathrm{Sat}}$ over the contribution from high
$T_{\mathrm{C}}$ ferromagnetic nanocrystals.}
\label{Fig:figANDhistANDfmt}
\end{figure*}

We begin by noting that the superparamagnetic-like component originates primarily from nanocrystals characterized by a relatively high magnitude of the spin ordering temperature, so that their magnetization $M_{\mathrm{N}}(T,H)$ can be regarded as temperature independent at very low temperatures. This means that a temperature dependence of the magnetization in this range comes from dilute Fe$^{3+}$ ions, whose properties in GaN have been extensively investigated previously.\cite{Pacuski:2008_PRL,Malguth:2008_pssb} 

\begin{figure}
   \begin{center}
        \includegraphics[width=0.97\columnwidth]{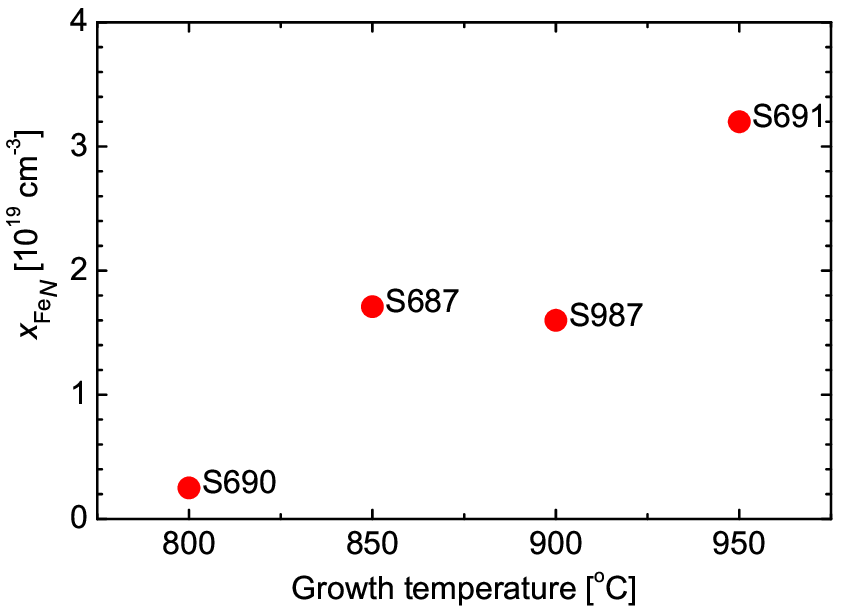}
   \end{center}
   \caption{(Color online) Estimated lower limit for the Fe concentration that precipitates in the form of various Fe-rich nanocrystals $x_{\mathrm{Fe}_N}$ as a function of the growth temperature.}
   \label{fig:figFMconc2}
\end{figure}

Accordingly, the concentration of these ions $x_{\mathrm{Fe}^{3+}}$ can be obtained by fitting $g\mu_BS$$x_{\mathrm{Fe}^{3+}}N_0\Delta B_{S}(\Delta T,H)$ to the difference between the experimental values of the magnetization measured at 1.85 and 5.0~K, where $\Delta B_{S}(\Delta T,H)$ is the difference of the corresponding paramagnetic Brillouin functions
$\Delta B_{S}(\Delta T,H)=B_{S}(1.85\, \mathrm{K},H)-B_{S}(5\, \mathrm{K},H)$. We consider the spin $S = 5/2$, the corresponding Land\'e factor $g=2.0$, and treat $x_{\mathrm{Fe}^{3+}}$ as the only fitting parameter.

The values established in this way are listed in Table~\ref{Tab:table1} for the studied samples and they are then employed to calculate the paramagnetic contribution at any temperature according to $M=g\mu_{\mathrm{B}}Sx_{\mathrm{Fe}^{3+}}B_{S}(T,H)$, which is then subtracted from the experimental data to obtain the magnitude of the magnetization $M_{\mathrm{N}}(T,H)$ coming from nanocrystals.

For the layers grown at $T_{\mathrm{g}} <$ 900$^{\circ}$C, $M_{\mathrm{N}}(T,H)$ saturates at all investigated temperatures for a magnetic field above $\sim10$~kOe, as evidenced in Fig.~\ref{Fig:figANDhistANDfmt}(a), pointing to a predominantly ferromagnetic order within the nanocrystals. The values of saturation magnetization $M_{\mathrm{N}}^{\mathrm{Sat}}$ obtained in this way when plotted \textit{vs.} temperature as in Fig.~\ref{Fig:figANDhistANDfmt}(b) allow us to assess the corresponding $T_{\mathrm{C}}$ from a fitting of the classical Brillouin function to the experimental points. Furthermore, assuming a value of the magnetic moment
2$\mu_{\mathrm{B}}$ per Fe, as in $\varepsilon$-Fe$_3$N,\cite{Bouchard:1974_JAP} we determine the concentration
of Fe ions $x_{\mathrm{Fe}_N}$ contributing to the Fe-rich nanocrystals, as shown in Table~I.

However, for the samples deposited at $T_{\mathrm{g}} \geq$ 900$^{\circ}$C the magnitude of $M_{\mathrm{N}}(H)$ saturates only at relatively high temperatures, namely around $T\gtrsim 150$~K, whereas at low temperatures it shows the sizable contribution of a slowly saturating component, as shown in Fig.~\ref{Fig:figANDhistANDfmt}(c) where magnetization data acquired at $1.85$~K for the layer S691 are reported.

This new contribution must arise from magnetically coupled objects with a spin arrangement other than ferromagnetic. According to the SXRD measurements previously discussed and summarized in Fig.~\ref{fig:SXRD} the most likely candidate is orthorombic $\zeta$-Fe$_2$N, antiferromagnetic below 9~K,\cite{Hinomura:1996_INC} or slowly saturating weakly ferromagnetic below 30~K.\cite{Nagamura:2004_STAM} In this case, in order to  establish $M_{\mathrm{N}}^{\mathrm{Sat}}$ we employ the Arrot plot method. The value of $M_{\mathrm{N}}^{\mathrm{Sat}}(T)$ determined in this way is reported in Fig.~\ref{Fig:figANDhistANDfmt}(d), and is seen to differ considerably from that of layers grown at lower $T_{\mathrm{g}}$. 

We are able to approximate the experimental values of $M_{\mathrm{N}}^{\mathrm{Sat}}(T)$ with a single Brillouin function only for $T\gtrsim150$~K (dashed line in Fig.~\ref{Fig:figANDhistANDfmt}(d)). This points to a lower value of $T_{\mathrm{C}}\cong 430$~K, indicating a shift of the chemical composition of $\varepsilon$-Fe$_{3-x}$N from Fe$_3$N ($x\cong 0$) for $T_{\mathrm{g}} <$ 900$^{\circ}$C to at most Fe$_{2.6}$N ($x\cong 0.4$) for $T_{\mathrm{g}}\geq$ 900$^{\circ}$C, as $T_{\mathrm{C}}$ of $\varepsilon$-Fe$_{3-x}$N decreases with increasing nitrogen content.\cite{Bouchard:1974_JAP} 

Moreover, the gradually increasing values of $M_{\mathrm{N}}^{\mathrm{Sat}}(T)$ below $T\lesssim 150$~K, marked as the hatched area in Fig.~\ref{Fig:figANDhistANDfmt}(d), indicate the presence of even more diluted $\varepsilon$-Fe$_{3-x}$N nanocrystals with $x$ ranging from 0.5 to 1 and with a wide spectrum of $T_{\mathrm{C}}$. Importantly, since $\varepsilon$-Fe$_{3-x}$N preserves its crystallographic structure and the changes of the lattice parameters are minor in the whole range $0\leq x \leq 1$, all various $\varepsilon$-Fe$_{3-x}$N nanocrystals contribute to the same diffraction peak in the SXRD spectrum, and are detected there as a single compound.

We note that the presence of either $\varepsilon$-Fe$_3$N or $\zeta$-Fe$_2$N, characterized by a low spin ordering temperature, does not hinder    the determination of the $x_{\mathrm{Fe}^{3+}}$ values, as both compounds have a rather low magnetic moment of $0.1\mu_B$ per Fe atom. Accordingly, the resulting variation of their magnetization is small comparing to the changes of the Fe$^{3+}$ paramagnetic signal at low temperatures.

The procedure exemplified above allows us to establish the lower limit for the Fe concentration that precipitates in the form of various Fe-rich nanocrystals ($x_{\mathrm{Fe}_N}$), and that is determined by the magnitude of $M_{\mathrm{N}}^{\mathrm{Sat}}$ at low temperatures. We again assume $2\mu_{\mathrm{B}}$ per  Fe atom, as in the dominant $\varepsilon$-Fe$_3$N. These values are collected in Table~\ref{Tab:table1} and plotted as the function of $T_{\mathrm{g}}$ in Fig.~\ref{fig:figFMconc2}. We see, that $x_{\mathrm{Fe}_N}$ consistently increases with $T_{\mathrm{g}}$, and that the growth temperature plays a more crucial role than the Fe-precursor flow rate~\cite{Bonanni:2007_PRB} in establishing the total value of $x_{\mathrm{Fe}_N}$.

\begin{figure} [h]
   \begin{center}
        \includegraphics[width=0.97\columnwidth]{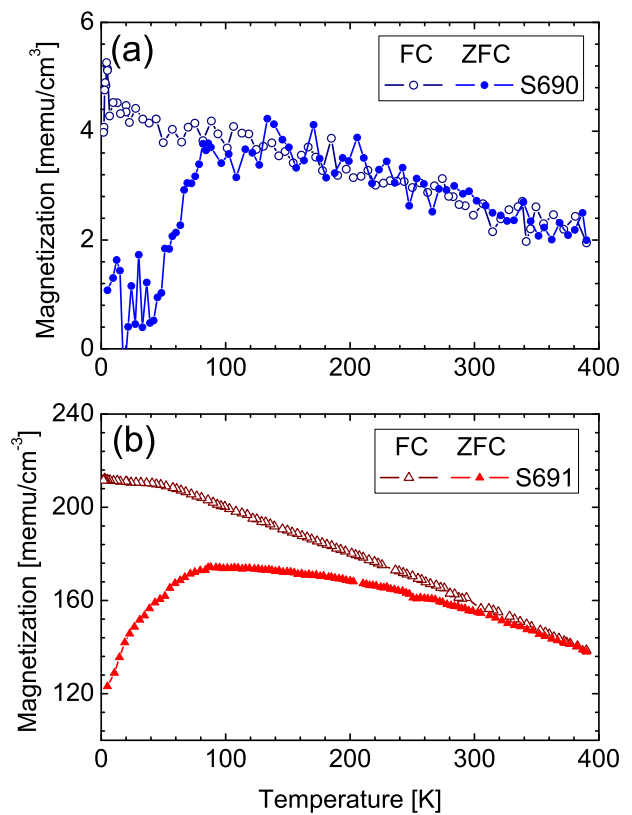}
   \end{center}
   \caption{(Color online) ZFC and FC curves measured in applied magnetic field of $200$~Oe for samples grown at (a) 800$^{\circ}$C and (b) 950$^{\circ}$C.}
   \label{fig:figFCZFC}
\end{figure}

Finally, measurements of FC and ZFC magnetization confirm the superparamagnetic-like behavior of $M_{\mathrm{N}}(T,H)$, as reported in Fig.~\ref{fig:figFCZFC}. As seen in Fig.~\ref{fig:figFCZFC}(a), the layer grown at the lowest temperature (S690) shows a minimal spread in $T_{\mathrm{B}}$, having its maximum at $T_{\mathrm{B}} < 100$~K, and accordingly a non-zero coercivity is evident only at low temperatures. 

In contrast, for most of the studied layers a broad maximum on the ZFC curve, exemplified in Fig.~\ref{fig:figFCZFC}(b), indicates a wide spread of blocking temperatures ($T_{\mathrm{B}}$) -- reaching RT -- and consequently a broad distribution in the volume of the nanocrystals. These high values of $T_{\mathrm{B}}$ are responsible for the existence of the open hysteresis in the $M(H)$ curves seen in Figs.~\ref{Fig:figANDhistANDfmt}(a),(c) and thus of a non-zero coercivity. This observation again points to the growth temperature as to the key factor in the determination of the crystallographic structure, size and chemical composition of the Fe-rich nanocrystals.

\section{Conclusions}
The previous\cite{Bonanni:2007_PRB,Bonanni:2008_PRL,Rovezzi:2009_PRB} and
present studies allow us to draw a number of conclusions concerning the
incorporation of Fe into GaN and about the resulting magnetic properties,
expected to be generic for a broad class of magnetically doped
semiconductors and oxides. These materials show magnetization consisting
typically of two components: i) a paramagnetic contribution appearing at
low temperatures and with characteristics typical for dilute magnetic
semiconductors containing weakly interacting randomly distributed magnetic
moments; ii) a puzzling ferromagnetic-like component persisting up to
above RT but with a value of remanence much smaller
than the magnitude of saturation magnetization.

According to SQUID and electron paramagnetic resonance~\cite{Bonanni:2007_PRB} measurements on
(Ga,Fe)N, the concentration of Ga substituting the randomly distributed
Fe$^{3+}$ ions increases with the iron precursor flow rate
reaching typically the value of 0.1\%. Our results imply that the
magnitude of the paramagnetic response and, hence, the density of dilute
Fe cations, does not virtually depend on the growth temperature. However,
the incorporation of Fe can be enlarged by co-doping with Si donors,
shifting the solubility limit to higher Fe concentrations.\cite{Bonanni:2008_PRL}

The presence of ferromagnetic-like features can be consistently
interpreted in terms of crystallographic and/or chemical phase separations
into nanoscale regions containing a large density of the magnetic
constituent. Our extensive SQUID, SXRD, TEM, EXAFS, and XANES measurements
of MOVPE-grown (Ga,Fe)N indicate that at the lowest growth temperature
$T_{\mathrm{g}} = 800^\circ$C, a large majority of the Fe ions occupy random
Ga-substitutional positions. However, in films grown at higher
temperatures,  $850 \leq T_{\mathrm{g}} \leq 950^\circ$C, a considerable
variety of Fe-rich nanocrystals is formed, differing in the Fe to N ratio.
In samples deposited at  the low end of the $T_{\mathrm{g}}$ range,
we observe mostly $\varepsilon-$Fe$_3$N precipitates but also inclusions of
elemental $\alpha$- and $\gamma$-Fe as well as of $\gamma'$-Fe$_4$N. In
all these materials $T_{\mathrm{C}}$ is well above RT,
so that the presence of the  corresponding nanocrystals explains the
robust superparamagnetic behavior of (Ga,Fe)N grown at $T_{\mathrm{g}}
\geq 850^\circ$C.

With the increase of the growth temperature nanocrystals of
$\zeta$-Fe$_2$N form and owing to antiferromagnetic interactions specific
to this compound, the magnetization acquires a component linear in the
magnetic field. This magnetic response 
has been previously observed and assigned to the Van Vleck paramagnetism
of isolated Fe$^{2+}$ ions. In view of the present findings, however, its
interpretation in terms of antiferromagnetically coupled spins in nitrogen
rich Fe$_x$N ($x \le 2$) nanocrystals seems more grounded.

The total amount of Fe ions contributing to the formation of the Fe-rich
nanocrystals is found to increase with the lowering of the growth rate
and/or with the raising of the growth temperature. At the same time,
however, the size of individual nanocrystals appears not to vary with
the growth parameters. Furthermore, annealing of (Ga,Fe)N containing only
diluted Fe cations does not result in a crystallographic phase separation.
Altogether, our findings indicate that the aggregation of Fe ions occurs
by nucleation at the growth front and is kinetically limited.
Moreover, according to the TEM results presented here, the spatial distribution of
nanocrystals is highly non-random. They tend to reside in two-dimensional
planes, particularly at the film surface and at the interface between the GaN buffer and the nominally Fe-doped layer. 

As a whole, these findings constitute a significant step
on the way to control the chemistry and local structure of
semiconductor/ferromagnetic metal nanocomposites.\\

\begin{acknowledgments}
The work was supported by the European Research Council through the FunDMS Advanced Grant within the "Ideas" 7th Framework Programme of the EC, and by the Austrian Fonds zur {F\"{o}rderung} der wissenschaftlichen Forschung -- FWF (P18942, P20065 and N107-NAN). We acknowledge the technical staff at the Rossendorfer Beamline (BM20) of the ESRF, and in particular C. B\"{a}htz and N. Jeutter for their valuable assistance. We also thank R. Jakie{\l}a for performing SIMS measurements.
\end{acknowledgments}



\end{document}